\pdfoutput=1

\documentclass[11pt]{article}

\usepackage[final]{acl}

\usepackage{times}
\usepackage{latexsym}

\usepackage[T1]{fontenc}

\usepackage[utf8]{inputenc}

\usepackage{microtype}

\usepackage{inconsolata}

\usepackage{graphicx}

\usepackage{amsmath}
\usepackage{breqn}
\usepackage{amsfonts}
\usepackage{dsfont}
\usepackage{multirow,makecell} 
\usepackage{booktabs}
\usepackage{siunitx}

\title{CIDRe: A Reference-Free Multi-Aspect Criterion for Code Comment Quality Measurement}

\author{
 \textbf{Maria Dziuba\textsuperscript{1,2}},
 \textbf{Valentin Malykh\textsuperscript{1,3}},
\\
 \textsuperscript{1}MTS AI,
 \textsuperscript{2}ITMO University,
 \textsuperscript{3}IITU University,
\\
 \texttt{{dziuba.maria}@niuitmo.ru}
 \\
 \texttt{valentin.malykh@phystech.edu}
}

\begin{document}
\maketitle
\begin{abstract}
Effective generation of structured code comments requires robust quality metrics for dataset curation, yet existing approaches (SIDE, MIDQ, STASIS) suffer from limited code-comment analysis. We propose CIDRe, a language-agnostic reference-free quality criterion combining four synergistic aspects: (1)~relevance (code-comment semantic alignment), (2)~informativeness (functional coverage), (3)~completeness (presence of all structure sections), and (4)~description length (detail sufficiency). We validate our criterion on a manually annotated dataset. Experiments demonstrate CIDRe’s superiority over existing metrics, achieving improvement in cross-entropy evaluation. When applied to filter comments, the models finetuned on CIDRe-filtered data show statistically significant quality gains in GPT-4o-mini assessments.
\end{abstract}

\section{Introduction}

Structured comments in docstring format, containing detailed descriptions of functionality, parameters, return values, exceptions, and use cases, play a key role in maintaining the codebase: they not only speed up developers' understanding of the code, but also allow them to automatically generate documentation (for example, in HTML format). %
Automatic comment generation can significantly ease the time-consuming task of writing comments and regularly updating them, which is necessary due to constant changes in the code base.

The work of Shi et al. \cite{shi2022we} demonstrates that the use of rule-based filters to delete low-quality comments greatly improves the generation results. Nevertheless, such simple filtering approaches are not enough, because they disregard the code-comment semantic similarity. Researchers attempted to use the SIDE metric~\cite{mastropaolo2024evaluating}, which includes Russian language support, to filter English training data \cite{vitale2025optimizing}. However, experiments revealed that even halving the dataset size had minimal impact on summary quality and model accuracy. This indicates the need to explore alternative quality metrics for optimizing code summarization datasets.

There was a trial to apply SIDE \cite{mastropaolo2024evaluating} metric to filter the comments~\cite{vitale2025optimizing}, but the results show that even halving the training set sizes does not significantly affect the model’s ability to generate summaries. However, when comparing the most restrictive selection strategy with a simpler one that randomly selects the training instances, the authors observe that the resulting accuracy of the model also does not change. This result suggests that different quality attributes should be explored for optimizing code summarization datasets.

Apart from SIDE, there are several different metrics, such as MIDQ~\cite{scalabrino2017automatically}, STASIS~\cite{li2006sentence}, or CoCC~\cite{huang2025your}, which have their own drawbacks: MIDQ, although considers comment structure, uses a Flesch index \cite{flesch1979write} designed for literary texts, which is incorrect for technical documentation full with terms. STASIS, based on WordNet \cite{miller1995wordnet} for English, is inapplicable for Russian due to the lack of an equivalent lexical base, and also does not take into account the different informative content of terms (for example, abbreviations like "id" or "ctx"). CoCC is a trained from scratch skip-gram word2vec~\cite{mikolov2013efficient} for code-comment consistency detection, which also does not support Russian language.

The problem of evaluating the quality of comments is compounded by the limitations of existing metrics. Text-based reference approaches such as BLEU or ROUGE-L depend on the quality of the reference data, which may be incomplete or contain errors. These metrics also do not take into account the semantic equivalence of alternative formulations, artificially underestimating the assessment of correct comments that differ from the standard.

In this paper, we propose a new criterion called CIDRe for the quality of structured comments that eliminates dependence on reference data, enables dataset filtration and evaluates several aspects of quality at the same time. We evaluate CIDRe on StRuCom~\cite{strucom}, which is the only existing dataset with strict structural filtering of code comments.

\textbf{Our contributions:}  
\begin{enumerate}
\item \textbf{Quality criterion for structured comments.} CIDRe is reference-free metric that combines four complementary quality components and outperforms existing approaches in cross-entropy evaluation.

\item \textbf{Validation dataset.} We manually annotated 840 comments from StRuCom for binary classification (good/bad), creating the first training/evaluation dataset for quality assessment criteria in this domain.

\item \textbf{Validation through finetuning.} Filtering StRuCom dataset with our criterion improved generation quality (in Side-by-Side evaluation via \texttt{gpt4-o-mini}), confirming the practical value of our approach.

\end{enumerate}

\section{Related Work}
\subsection{Datasets}
Existing code-to-text datasets predominantly target English content. The Stack~\cite{Kocetkov2022TheS3} aggregates multilingual code (658 languages) but lacks task-specific annotations for supervised fine-tuning. The Vault~\cite{Mnh2023TheVA}, derived from The Stack, contains 43M English code-text pairs, yet structured documentation remains scarce due to an abundance of brief functions. CodeSearchNet~\cite{Husain2019CodeSearchNetCE} focuses on code search, limiting text descriptions to introductory documentation paragraphs. MCoNaLa~\cite{Wang2022MCoNaLaAB} offers minimal multilingual support (345 Russian examples) but is constrained to simple "how-to" Python snippets. StRuCom \cite{strucom} addresses the Russian documentation gap with 153K human-written and synthetic code-comment pairs across Python, Java, JavaScript, C\#, and Go, maintaining language-specific terminology and docstring conventions.

\subsection{Models}
While proprietary large language models (LLMs), such as GPT-4\footnote{\url{https://openai.com/index/gpt-4/}}, are excellent at generating code documentation, their proprietary nature poses a challenge for enterprise adoption. In contrast, open-source alternatives such as DeepSeek-Coder and Qwen2.5-Coder offer a balance between performance and deployability, although they may underperform on Russian documentation due to their training on English corpora only.
Recent research~\cite{strucom} has introduced models trained on the StRuCom dataset, which achieves baseline performance on Russian code commenting tasks. However, given the inherent noise in the dataset, arising from uncurated human-generated and synthetic data, there is potential for accuracy improvements through quality filtering, a direction that remains to be explored in the literature.

\subsection{Embedding Models for Code}
Modern embedding models bridge the gap between code and natural language through semantic alignment. CodeSage~\cite{zhang2024code} utilizes bidirectional transformers with scalable architectures (130M–1.3B parameters) to align code-text representations through contrastive learning, though its English-centric training limits Russian adaptability. CodeXEmbed~\cite{liu2024codexembed} employs a unified multilingual framework (400M–7B parameters) for cross-modal retrieval across 12 languages, achieving state-of-the-art benchmarks at the cost of high computational overhead. Both models highlight the necessity of linguistic adaptation for non-English documentation tasks.

\subsection{Metrics for Comment Quality}
Existing metrics exhibit language and domain limitations. MIDQ~\cite{scalabrino2017automatically} combines JavaDoc structure analysis with Flesch readability scores, though its reliance on literary readability and Java-specific design hinders cross-lingual applicability. STASIS~\cite{li2006sentence} measures code-comment similarity via WordNet synsets, computing term distances in the WordNet hierarchy, but suffers from English language bias, uniform term weighting, and lack of Russian support. CoCC~\cite{huang2025your} detects inconsistencies through code-text embeddings but remains English-centric. SIDE~\cite{mastropaolo2024evaluating} introduces reference-free coherence evaluation via contrastive learning with MPNet \cite{song2020mpnet} (12-layer BERT \cite{devlin2019bert} architecture), yet empirical studies show its failure to improve model performance when filtering TL-CodeSum \cite{hu2018summarizing} and Funcom \cite{leclair2019recommendations} datasets – highlighting the need for multi-aspect quality criteria.

\section{CIDRe Criterion}
 
\begin{figure}[tbh!]
    \centering
    \includegraphics[width=1\linewidth]{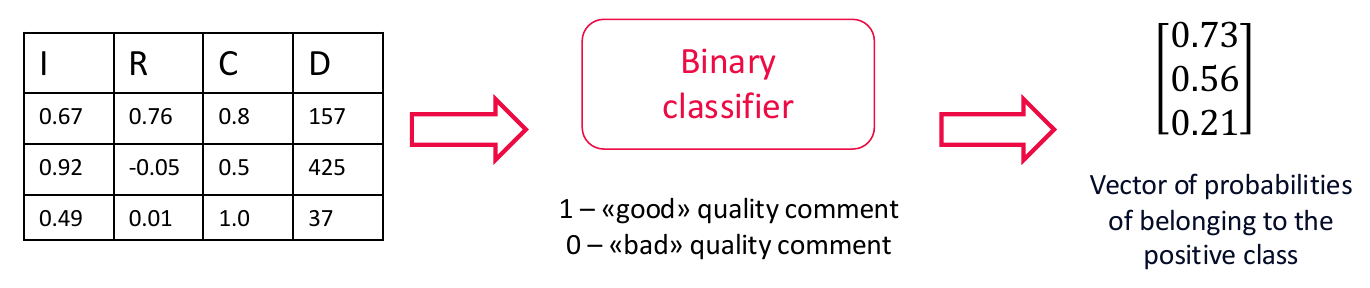}
    \caption{Criteria pipeline. The proposed four quality measures are considered for the proposed list of <<code-comment>> pairs, the resulting matrix is an input to a binary classifier (class $1$ - <<good>> comments, class $0$ - <<bad>> comments), which in turn outputs a vector of probabilities that comments belong to first class. Thus, the values of our criterion are real numbers belonging to the segment $[0, 1]$.}
    \label{fig:pipeline}
\end{figure}

Our criterion is a combination of four key components: \textbf{C}ompleteness, \textbf{I}nformativeness, length of the text \textbf{D}escription, and \textbf{Re}levance. The general pipeline of the criterion is shown in Figure \ref{fig:pipeline}. 

\subsection{Completeness}
\textbf{Completeness} is a measure of the documentation's coverage of code elements. The documented elements of a function are its parameters, the exceptions thrown from it, and its return value. Our definition of completeness is inspired by the Documented Iterms Ratio (DIR) proposed by the authors of the MIDQ metric. But since MIDQ is defined only for JavaDoc, we have defined it for the other $4$ programming languages. The details of completeness calculation are placed in Appendix~\ref{sec:completeness_calculation}.

\subsection{Informativeness}  
\textbf{Informativeness} measures the extent to which information from a function’s source code is captured in its corresponding comment. This metric is grounded in the assumption that competent programmers assign semantically meaningful names to code identifiers, enabling critical insights into the code’s functionality to be derived directly from these names.  

A \textbf{term} is defined as a word contained within an identifier. A single identifier may comprise \(N\) words, thereby containing \(N\) terms.  

The measure of informativeness is inspired by STASIS. The main difference between informativeness and STASIS is the consideration of the weight of terms. The idea is as follows: the terms in the code have different meanings for understanding its functionality. Therefore, it is necessary to weigh the terms by importance, for this we use the mechanism of self-attention in transformers. The details about informativeness calculation are placed in Appendix~\ref{sec:inf_models}.

\subsection{Description Length}

We measure comment length in characters, hypothesizing that detailed textual explanations before key sections (parameters, returns, etc.) improve system-wide context understanding by linking functions to broader architectural goals, reduce cognitive load through self-contained explanations of complex logic, and preserve decision history across code iterations. Furthermore, comprehensive comments mitigate knowledge gaps in collaborative environments by explicitly documenting assumptions and edge cases. 

\subsection{Relevance}  
\textbf{Relevance} quantifies the degree of semantic alignment between generated code comments and the corresponding source code. The methodology is developed based on a SIDE metric. We employed a Triplet Loss function \cite{schroff2015facenet} to finetune CodeSage-small-v2 model, which choice is justified by its support for diverse languages and its ability to generate semantically rich embeddings for code-text matching. For details about finetuning see Appendix~\ref{sec:relevance_model_finetuning}. 

\section{Verification of the  Components}

Tab.~\ref{tab:stat_significance} presents statistical analysis via Mann-Whitney U-test, which confirms significant differences (p < 0.05) between <<good>> and <<bad>> comments across all four components: informativeness, relevance, structural completeness, and description length. The observed patterns align with human judgments — high-quality comments systematically exhibit richer contextual details, adhere to documentation standards, and minimize redundant information. This empirical validation justifies the components' inclusion in the final quality criterion.

\begin{table}[tbh!]
    \centering
    \small
    \setlength{\tabcolsep}{4pt} 
    \begin{tabular}{@{}ccccr@{}}
        \toprule
        Metric & \multicolumn{2}{c}{Comment Quality} & p-value \\ 
        \cmidrule(lr){2-3}
        & Bad & Good & \\ 
        \midrule
        C & $0.85 \pm 0.20$ & $1.00 \pm 0.03$ & \textbf{$1.3 \times 10^{-41}$} \\ 
        I & $0.17 \pm 0.06$ & $0.26 \pm 0.08$ & \textbf{$9.2 \times 10^{-52}$} \\ 
        
        D & $342.7 \pm 153.2$ & $470.2 \pm 166.9$ & \textbf{$5.0 \times 10^{-36}$} \\ 
        R & $0.76 \pm 0.11$ & $0.84 \pm 0.08$ & \textbf{$1.6 \times 10^{-24}$} \\ 
        \bottomrule
    \end{tabular}
    \caption{Comparison of quality criterion components between comment groups by Mann-Whitney test.}
    \label{tab:stat_significance}
\end{table}

\section{Metric Comparison}
We compare our proposed criterion with MIDQ~\cite{scalabrino2017automatically}, which relies on JavaDoc structure analysis and Flesch readability scores, and SIDE~\cite{mastropaolo2024evaluating}, which is reference-free coherence evaluation with MPNet. 
We evaluate CIDRe against the existing metrics using an independent test set of 100 code comments with cross-entropy, which penalizes both classification errors and probabilistic miscalibrations — notably harsh on overconfident incorrect predictions. As can be seen in Tab.~\ref{tab:cross_entropy}, the experiments demonstrate our SVM-based approach's superiority in probability calibration over ensemble and linear methods, which suffer from error accumulation and nonlinearity handling limitations respectively. Traditional documentation metrics (SIDE, MIDQ) underperform in confidence-sensitive scenarios, validating the need for specialized criteria in borderline case analysis.

\begin{table}
    \centering
    \begin{tabular}{lc}
        \hline
         Metric & CE \\
         \hline
         CIDRe (SVM) & \textbf{1.35} \\
         CIDRe (LightGBM) & $5.66$ \\
         CIDRe (Logistic Regression) & $16.01$ \\
         SIDE~\cite{mastropaolo2024evaluating} & $5.32$ \\
         MIDQ~\cite{scalabrino2017automatically} & $7.55$ \\
         \hline
    \end{tabular}
    \caption{Comparison of the developed quality criterion based on three different models and two baselines by cross-entropy (CE) with existing metrics for the quality of code comments.}
    \label{tab:cross_entropy}
\end{table}

\section{Ablation study}  
We analyze the contribution of individual features (informativeness, relevance, completeness, description length) to our quality criterion using SVM as presented in Tab.~\ref{tab:ablation_study}. Progressive exclusion experiments reveal that the full feature set achieves optimal performance, with any partial combination degrading results monotonically. The synergy between features is evident — omitting any single component disproportionately reduces model effectiveness, validating our four-dimensional design.

\begin{table}[tbh!]
    \centering
    \small
    \setlength{\tabcolsep}{3pt}
    \begin{tabular}{@{}*{4}{cc}@{}}
        \toprule
        \multicolumn{2}{c}{1 Feature} & \multicolumn{2}{c}{2 Features} & \multicolumn{2}{c}{3 Features} & \multicolumn{2}{c}{4 Features} \\
        \cmidrule(lr){1-2} \cmidrule(lr){3-4} \cmidrule(lr){5-6} \cmidrule(lr){7-8}
        F & F1 & F & F1 & F & F1 & F & F1 \\
        \midrule
        I    & 0.747 & I,D   & 0.840 & I,R,C & 0.945 & I,R,C,D & 0.994 \\
        R    & 0.719 & I,C   & 0.839 & I,R,D & 0.945 \\
        D    & 0.724 & R,C   & 0.802 & I,C,D & 0.950 \\
        C    & 0.779 & R,D   & 0.844 & R,C,D & 0.890 \\
        I,R  & 0.860 & C,D   & 0.816 \\
        \bottomrule
    \end{tabular}
    \caption{Ablation study for SVM model (F1-score). Key: I-informativeness, R-relevance, C-completeness, D-description length. Best result (0.994 F1) with all features.}
    \label{tab:ablation_study}
\end{table}

\section{Side-by-Side Evaluation}  

The comparison was performed with GitHub Copilot on a test subset of StRuCom using LLM-as-judge method (the judge is GPT-4o-mini). We finetuned Qwen2.5-Coder models of different size Qwen2.5-Coder (0.5B-7B).

Experiments demonstrate our criterion's universal effectiveness: data filtering improves model metrics across architectures and languages by removing noise while preserving semantically critical comment patterns. More details are presented in Appendix~\ref{sec:sbs}. 

\section{Conclusion}
We proposed a reference-free quality criterion for code comments, evaluating semantics alignment, functionality coverage, parameter/exception completeness, and text length. Validated on 840 manually annotated comments, our approach outperformed existing metrics and improved generation quality when filtering the StRuCom dataset, as confirmed by GPT-4o-mini based evaluations. Future work will extend the criterion to other programming and natural languages, enhancing its applicability for multilingual documentation systems. 

\section{Limitations}
While our criterion advances quality assessment for structured comments, it may not comprehensively capture all quality dimensions, such as stylistic consistency or domain-specific terminology appropriateness. The metric is specifically optimized for Russian-language documentation and structured formats (e.g., docstrings), potentially assigning disproportionately low scores to comments in other languages or free-form styles due to linguistic and structural biases. 

\bibliography{custom}

\appendix

\section{Completeness Calculation}
\label{sec:completeness_calculation}
\begin{figure*}[tbh!]
    \centering
    \includegraphics[width=1.0\linewidth]{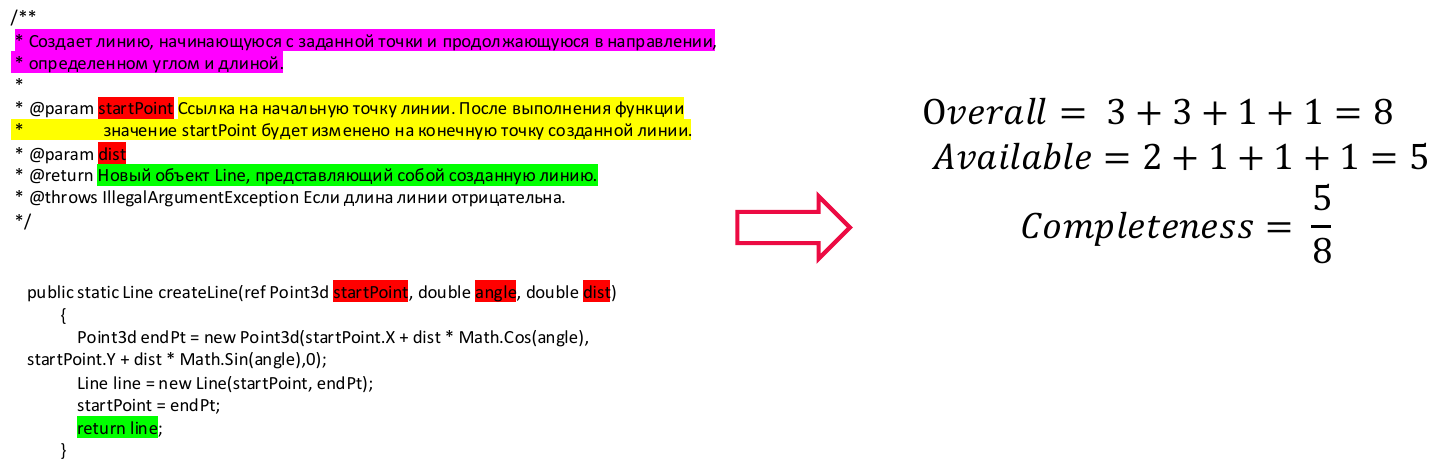}
    \caption{An example of completeness calculation}
    \label{fig:completeness_example}
\end{figure*}

Let's consider the formulas for calculating completeness. First, the number of documented elements is calculated using the formula \ref{eq:overall_java} or \ref{eq:overall_python} (depending on the programming language). Next, the number of found documentation elements in this comment is calculated using the formula \ref{eq:available_java} or \ref{eq:available_python}. Completeness for Go is considered separate from other programming languages due to the very simple comment format, and can be either $0$ or $1$, see the formula \ref{eq:completeness_go}. Completeness for other languages is calculated as the ratio of the number of documented elements to the number of found elements with documentation, see the formula \ref{eq:completeness_java}. An example of calculating completeness is shown in the figure \ref{fig:completeness_example}.

\begin{multline}
    Overall_{Java/C\#} = 2 \cdot (|E_{code}|) + int(R_{code}) + \\ 2 \cdot (|P_{code}|) +int(count\_description)
\label{eq:overall_java}
\end{multline}

\begin{multline}
    Overall_{Python/JS} = 2 \cdot (|E_{code}|) + 2 \cdot int(R_{code}) + \\ 3 \cdot (|P_{code}|) + int(count\_description)
\label{eq:overall_python}
\end{multline}

\begin{multline}
    Available_{Java/C\#} = (|E_{code} \cap E_{comment}| + \\ E_{n\_descriptions}) + int(R_{code} \land R_{comment})  + \\ (|P_{code} \cap P_{comment}| + P_{n\_descriptions}) + \\ int(has\_description \land count\_description)
\label{eq:available_java}
\end{multline}

\begin{multline}
    Available_{Python/JS} = (|E_{code} \cap E_{comment}| \\  + E_{n\_descriptions}) + (int(R_{code} \land R_{comment}) \\ \cdot (int(R_{type\_name} + int(R_{description}))))  + \\ (|P_{code} \cap P_{comment}| + P_{n\_descriptions} + \\ P_{n\_type\_names}) + int(has\_description \land \\ count\_description)
\label{eq:available_python}
\end{multline}

\begin{dmath}
    Completeness_{Go} = 
    \begin{cases}
    1, comment.starts\_with(function\_name) \\
    0, \text{otherwise}
    \end{cases}
\label{eq:completeness_go}
\end{dmath}

\begin{dmath}
    Completeness_{Java/JS/C\#/Python} = \frac{Available}{Overall}
\label{eq:completeness_java}
\end{dmath}

where $E_{code}, E_{comment}$ are the sets of \textbf{exceptions} in the code and comments, respectively, $E_{n\_descriptions} \in \mathbb{N}$ is the number of text descriptions for exceptions $e \in E_{code}$, $P_{code}, P_{comment}$ - sets of \textbf{parameters} of the function in the code and comments, respectively, $P_{n\_descriptions} \in \mathbb{N}$ - number of text descriptions for parameters $p \in P_{code}$, $R_{code} \in \{0, 1\}$ $R_{comment}, R_{type\_name}, R_{description} \in \{0, 1\}$ are Boolean variables that indicate whether a comment contains a return value, its type, and a textual description, $has\_description \in \{0, 1\}$ - boolean variable indicating whether there is a description in the comment (before the list of structural tags), $count\_description \in \{0, 1\}$ Boolean variable indicating whether the text description should be counted is $0$ if there is at least one structural tag in the comment.

\section{Models for Informativeness Weights}
\label{sec:inf_models}
\begin{figure*}[tbh!]
    \centering
    \includegraphics[width=1\linewidth]{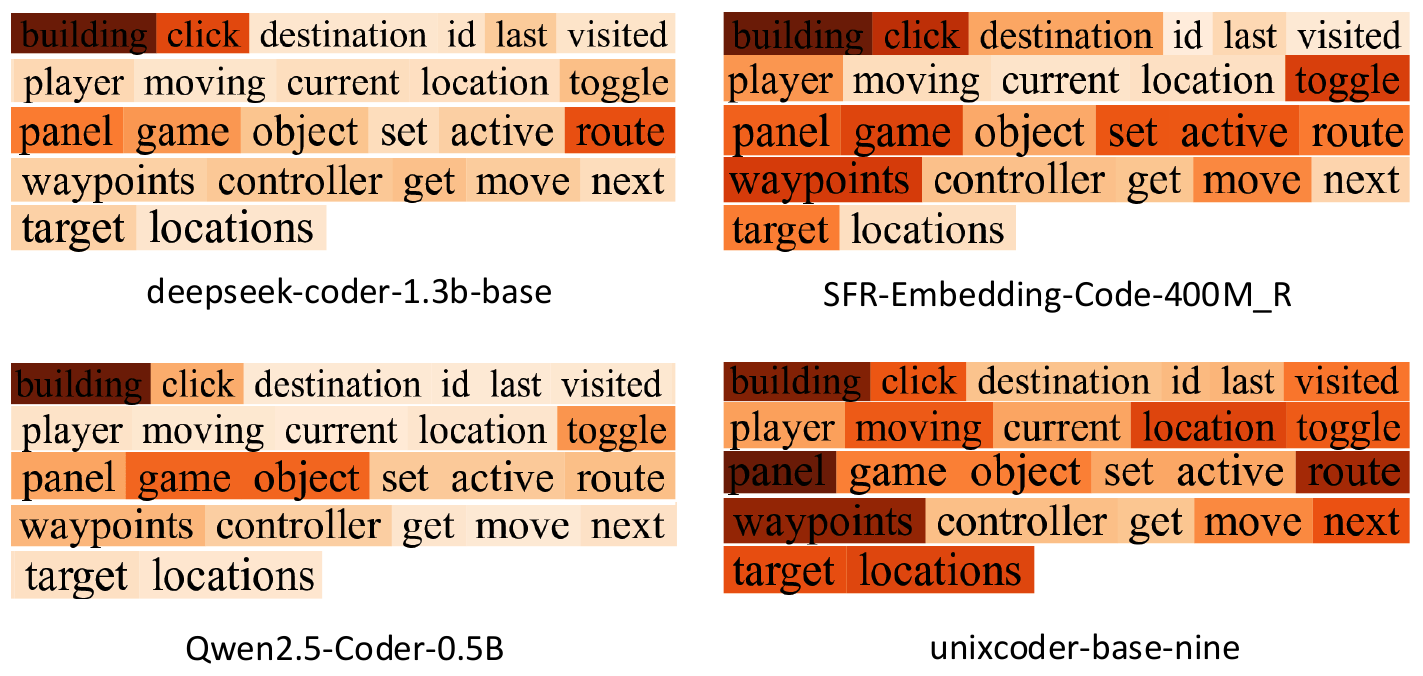}
    \caption{Visualization of the weighting of terms by importance using different models}
    \label{fig:inf_model_comparison}
\end{figure*}

This study conducted a comparative analysis of four deep learning models computing attention weights: \textit{Qwen2.5-Coder-0.5B}, \textit{deepseek-coder-1.3b-base}, \textit{unixcoder-base-nine}, and \textit{SFR-Embedding-Code-400M\_R} (CodeXEmbed). Figure \ref{fig:inf_model_comparison} visualizes the example of term importance distributions for each model. Based on the evaluation, \textit{SFR-Embedding-Code-400M\_R} was selected as the baseline architecture for informativeness computation. Key experimental findings include:  

\begin{enumerate}  
    \item The \textit{Qwen2.5-Coder} and \textit{deepseek-coder-1.3b-base} models exhibited excessive selectivity, disregarding semantically significant terms (e.g., the term \textit{player} in test cases received insufficient attention weights).  
    \item \textit{Unixcoder-base-nine} assigned disproportionately high weights to low-significance elements (such as \textit{id}), reducing analytical precision.  
    \item \textit{SFR-Embedding-Code-400M\_R} demonstrated balanced term weighting, maintaining feature relevance with moderate computational complexity (400 million parameters).  
\end{enumerate}  

Thus, the \textit{SFR-Embedding-Code-400M\_R} architecture achieves balanced highlighting of semantically significant terms, avoiding both underestimation of critical elements and overemphasis on trivial tokens. The model’s compact size (400 million parameters) enables high-speed data processing while preserving analytical accuracy. This combination of characteristics renders the selected approach optimal for automated code informativeness assessment tasks.

\section{Relevance Model Finetuning}
\label{sec:relevance_model_finetuning}
To adapt the model for evaluating the relevance of structured Russian-language comments, a specialized dataset was constructed from StRuCom. Positive examples (\(1500\) instances) include comments evenly distributed across five programming languages (\(300\) examples per language). Negative examples (\(3000\) instances) comprise semantically similar negatives selected via the \textit{mine\_hard\_negatives} algorithm from the Sentence Transformers library.  

The hard negative mining algorithm employs a cosine similarity metric to retrieve the top-\(3\) examples most similar to the anchor, with a similarity threshold \(>0\). The total training corpus comprises \(4500\) instances.  

Fine-tuning was conducted for one epoch on an NVIDIA A100 GPU using the AdamW optimizer (learning rate \(5 \times 10^{-5}\)) with a batch size of \(16\). Regularization included a warmup phase with a coefficient of \(0.1\) and a Triplet Loss margin parameter of \(0.3\). 

\section{Informativeness Calculation}
To compute informativeness, terms are first extracted from the source code. This is achieved by parsing the code into an abstract syntax tree (AST) using the Code-Text parser library\footnote{\url{https://github.com/FSoft-AI4Code/CodeText-parser/tree/main}}. The identifiers undergo lemmatization, deduplication, and filtering to exclude stop words and terms with a length of \(1\). The comment text is similarly preprocessed. Words in the comment are lemmatized and filtered to remove stop words, ensuring semantic alignment with the extracted code terms. 

For terms' weighting, the attention weights from the last layer of the model are used, then they are averaged over all heads, and then the sum of all attention weights for each token is calculated (relative to itself and relative to other tokens). The details about how we chose a model for informativeness weights is placed in Appendix \ref{sec:inf_models}

\begin{figure*}[tbh!]
    \centering
    \includegraphics[width=1\linewidth]{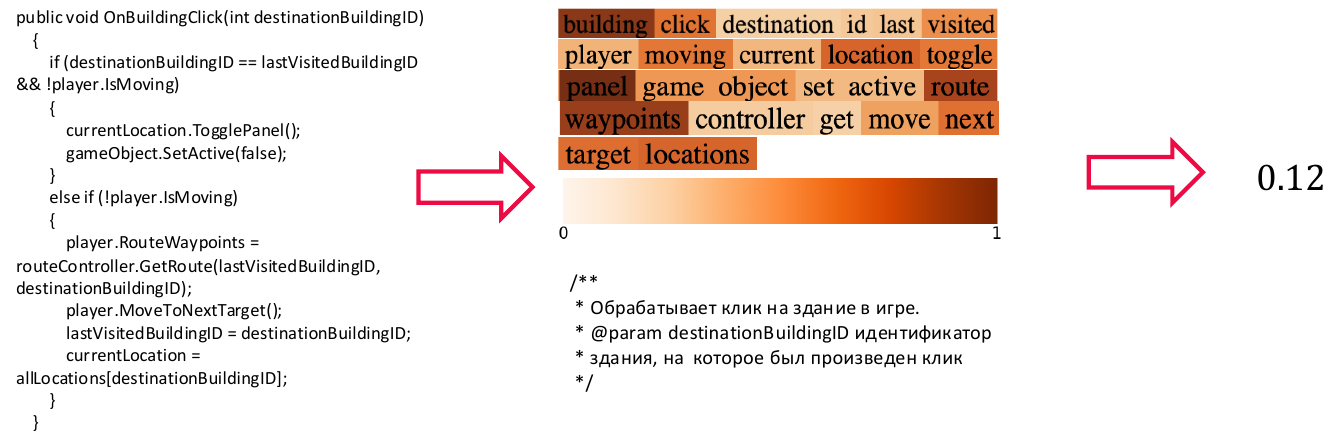}
    \caption{An example of informativeness calculation}
    \label{fig:informativeness_example}
\end{figure*}

Next, we calculate how many terms from the code are in this comment. For this, the cosine similarity between the embeddings of words in the comment and the embeddings of terms in the code is used. Multilingual embeddings ConceptNet Numberbatch \cite{speer2017conceptnet} are used as embeddings. If the cosine similarity between the embeddings of a term and a word is greater than the specified threshold ($0.5$), then the term is considered found. The final informativeness formula is presented in the formula \ref{eq:informativeness} and is equal to the ratio of the number of terms found to the total number of terms in the code. An example of informativeness calculation is shown in the figure \ref{fig:informativeness_example}

\begin{equation}
Informativeness = \frac{found\_terms}{all\_terms}
\label{eq:informativeness}
\end{equation}

\section{Side-by-side comparison}
\label{sec:sbs}
We adopt the LLM-as-a-judge paradigm \cite{zheng2023judging}, leveraging GPT-4's RLHF-aligned reasoning for automated pairwise comparisons. This approach replaces costly expert labeling while maintaining ~80\% human judgment consistency and providing interpretable rationales. To mitigate positional bias, responses are evaluated twice with reversed order, recording victories only for consistent outcomes. Our implementation introduces two tie types: <<win>> (both responses adequate) and <<lose>> (both inadequate), refining outcome granularity for semantically similar comments.

\begin{table*}[tbh!]
\resizebox{\textwidth}{!}{  
    \centering
    \begin{tabular}{cccccc}
    \hline
        \makecell{Experiment type} & Python & Java & Go & C\# & JavaScript\\
        \hline
        \multicolumn{6}{c}{Qwen2.5-Coder-7B-Instruct} \\ %
        \hline
        \makecell{w/o finetuning} & \textbf{48.0}/2.0/16.5/33.5 & \textbf{65.5}/6.0/1.0/27.5 & \textbf{43.5}/3.5/6.0/47.0 & \textbf{22.0}/2.0/3.0/74.0 & \textbf{44.0}/4.5/3.0/48.5 \\
        \hline
        \makecell{w/o filtration} & \textbf{45.0}/6.5/19.0/29.5 & \textbf{85.0}/4.0/0.5/10.5 & \textbf{61.0}/5.5/5.0/28.5 & \textbf{81.0}/3.5/2.0/13.5 & \textbf{71.0}/2.0/0.0/27.0 \\
        \hline
        \makecell{w filtration} & \textbf{55.0}/3.0/6.5/35.5 & \textbf{87.0}/1.5/1.0/10.5 & \textbf{63.5}/4.5/4.0/28.0 & \textbf{79.0}/6.0/2.5/12.5 & \textbf{74.5}/5.5/0.5/19.5 \\
        \hline
        \multicolumn{6}{c}{Qwen2.5-Coder-3B-Instruct} \\ %
        \hline
        \makecell{w/o finetuning} & \textbf{7.0}/0.0/16.5/76.5 & \textbf{24.0}/0.5/2.5/73.0 & \textbf{7.0}/0.5/4.5/88.0 & \textbf{7.0}/0.0/4.5/88.5 & \textbf{19.5}/0.5/5.0/75.0 \\
        \hline
        \makecell{w/o filtration} & \textbf{41.5}/4.5/21.0/33.0 & \textbf{81.5}/6.0/0.0/12.5 & \textbf{58.0}/3.5/4.5/34.0 & \textbf{82.0}/4.0/2.0/12.0 & \textbf{65.5}/5.0/0.5/29.0 \\
        \hline
        \makecell{w filtration} & \textbf{58.0}/1.5/6.5/34.0 & \textbf{85.5}/2.0/0.5/12.0 & \textbf{62.5}/2.0/5.5/30.0 & \textbf{79.0}/5.0/3.5/12.5 & \textbf{78.5}/1.5/0.5/19.5 \\
        \hline
        \multicolumn{6}{c}{Qwen2.5-Coder-1.5B-Instruct} \\ %
        \hline
        \makecell{w/o finetuning} & \textbf{18.5}/0.5/16.5/64.5 & \textbf{20.0}/1.0/3.5/75.5 & \textbf{9.5}/0.0/8.5/82.0 & \textbf{7.0}/0.5/2.0/90.5 & \textbf{13.5}/0.0/3.5/83.0\\
        \hline
        \makecell{w/o filtration} & \textbf{38.0}/2.5/26.0/33.5 & \textbf{78.0}/4.0/1.5/16.5 & \textbf{58.0}/4.5/6.5/31.0 & \textbf{73.0}/4.5/3.5/19.0 & \textbf{58.5}/4.5/1.0/36.0 \\
        \hline
        \makecell{w filtration} & \textbf{46.0}/1.0/6.0/47.0 & \textbf{81.0}/4.0/0.5/14.5 & \textbf{57.0}/4.5/3.5/35.0 & \textbf{71.5}/4.5/4.0/20.0 & \textbf{65.0}/3.0/1.5/30.5 \\
        \hline
        \multicolumn{6}{c}{Qwen2.5-Coder-0.5B-Instruct} \\ %
        \hline
        \makecell{w/o finetuning} & 36.0/2.0/25.0/37.0/ & \textbf{24.5}/0.5/4.5/70.5 & \textbf{12.5}/0.0/13.5/74.0 & \textbf{5.5}/0.5/5.0/89.0 & \textbf{8.5}/0.0/4.0/87.5\\
        \hline
        \makecell{w/o filtration} & \textbf{18.0}/1.0/22.0/59.0 & \textbf{60.0}/3.5/2.0/34.5 & \textbf{31.5}/2.0/5.0/61.5 & \textbf{53.5}/2.5/4.0/40.0 & \textbf{41.0}/1.5/2.0/55.5 \\
        \hline
        \makecell{w filtration} & \textbf{27.5}/2.0/3.5/67.0 & \textbf{59.5}/4.0/0.5/36.0 & \textbf{40.5}/4.0/3.0/52.5 & \textbf{54.5}/3.0/2.5/40.0 & \textbf{51.5}/2.0/0.5/51.5 \\
        \hline
    \end{tabular}
}

    \caption{The results of the Side-by-side evaluation with the GPT-4o-mini judge. The estimates are presented as: Model VS Copilot, win/win\_tie/lose\_tie/lose, which corresponds to the estimates of 10/11/00/01. The answers were evaluated twice with a change in their order to solve the problem of positional bias.}
    \label{tab:sbs}
\end{table*}

The following experiments were conducted: with a model without additional training; with additional training, but without filtering with our criterion; with additional training and filtering of data with our criterion. We removed data with predicted probability of belonging to the first class (high quality comments) < 0.5.

We finetuned \textit{Qwen2.5-Coder} using LoRA adapters \cite{hu2022lora} on 4×NVIDIA A100 GPUs with DeepSpeed’s ZeRO Stage~2 ~\cite{rajbhandari2020zero} optimization. Training spanned 5 epochs with 2K-token sequences, employing AdamW ~\cite{loshchilov2017decoupled} ($\text{lr}=1\times10^{-4}$ to $1\times10^{-5}$ cosine decay), 1\% linear warmup, and 0.1 weight decay.

The results are shown in Table \ref{tab:sbs}. The conducted experiments demonstrate that additional filtering of data according to the developed criterion consistently improves metrics for most languages and models, reducing the frequency of partial and obvious failures. This effect is consistent for both large and minimal architectures, which indicates the universality of the criterion — it effectively filters out noise examples, preserving semantically rich patterns that are critical for generating meaningful comments regardless of the scale of the model.

\end{document}